
\input phyzzx
\input tables
\sequentialequations

\overfullrule=0pt
\catcode`\@=11

\def \F{\phi}

\def \D{{\delta}}

\def\NP{{ Nucl. Phys.\ }}

\def\PL{{ Phys. Lett.\ }}
\def\PR{{ Phys. Rev.\ }}

\def\IJMP{{ Int. Jour. Mod. Phys.\ }}

\def\lc{light-cone}
\def\lcq{light-cone quantization}

\def\eqaligntwo#1{\null\,\vcenter{\openup\jot\m@th
\ialign{\strut\hfil
$\displaystyle{##}$&$\displaystyle{{}##}$&$\displaystyle{{}##}$\hfil
\crcr#1\crcr}}\,}
\catcode`\@=12

\REF\Stan{For a review, see S. J. Brodsky and H.-C. Pauli,
``Light-Cone Quantization of Quantum Chromodynamics,''
Lectures at the 30th Schladming Winter School in
Particle Physics, SLAC-PUB-5558 (1991).}
\REF\Brod{H.-C. Pauli and S. Brodsky, \PR {\bf D32} (1985) 1993, 2001;
\nextline
K. Hornbostel, S. Brodsky and H.-C. Pauli, \PR {\bf D41} (1990) 3814;
\nextline
for a good review, see K. Hornbostel, Ph.D. thesis, SLAC report
No. 333 (1988).}
\REF\Thorn{C.~B. Thorn, \PL {\bf 70B} (1977) 85; \PR {\bf D17} (1978)
1073; {\bf D19} (1979) 639; {\bf D20} (1979) 1435.}
\REF\Tang{A. C. Tang, S. J. Brodsky, and H. C. Pauli,
\PR {\bf D44} (1991) 1842;\nextline
D. Mustaki, S. Pinsky, J. Shigemitsu,
and K. Wilson, \PR {\bf D43} (1991) 3411;\nextline
A. Harindranath and R.~J. Perry, \PR {\bf D43} (1991) 492;\nextline
A. Harindranath, R.~J. Perry and J. Shigemitsu, \PR {\bf D46} (1992) 4580.}
\REF\DK{S. Dalley and I. R. Klebanov, \PR {\bf D47} (1993) 2517.}
\REF\GH{G. 't Hooft, \NP {\bf B72} (1974) 461.}
\REF\qcd{G. 't Hooft, \NP {\bf B75} (1974) 461.}
\REF\FS{S.-S. Shei and H.-S. Tsao, \NP {\bf B141} (1978) 445.}
\REF\FSmore{W. Bardeen and R. Pearson, \PR {\bf D14} (1976) 547;
\nextline
M.~B. Halpern and P. Senjanovi\'c, \PR {\bf D15} (1977) 1655;\nextline
T.~N. Tomaras, \NP {\bf B163} (1980) 79.}
\REF\BDK{G. Bhanot, K. Demeterfi and I.~R. Klebanov, ``(1+1)-dimensional
large-$N$ QCD coupled to adjoint fermions,'' PUPT-1413, IASSNS-HEP-93/42,
{\tt hep-th/9307111} (to appear in Phys. Rev. D).}
\REF\dkut{D. Kutasov, ``Two Dimensional QCD coupled to Adjoint
Matter and String Theory,'' EFI-93-30, {\tt hep-th/9306013}.}
\REF\BPR{W. A. Bardeen, R. B. Pearson and E. Rabinovici,
\PR {\bf D21} (1980) 1037.}
\REF\KS{I. R. Klebanov and L. Susskind, \NP {\bf B309} (1988) 175.}
\REF\KKS{M. Karliner, I.~R. Klebanov and L. Susskind,
\IJMP {\bf A3} (1988) 1981.}
\REF\ext{R. Bulirsch and J. Stoer,  Numer. Math. {\bf 6} (1964) 413;
\nextline
M. Henkel and G. Sch\"utz,  J. Phys. A {\bf 21} (1988) 2617.}
\REF\kog{I. Kogan, ``Hot Gauge Theories and $\IZ_N$ Phases,'' PUPT-1415.}
\REF\DG{D. J. Gross, \NP {\bf B400} (1993) 161;\nextline
D. J. Gross and W. Taylor, \NP {\bf B400} (1993) 181;
{\bf B403} (1993) 395.}

\font\cmss=cmss10 \font\cmsss=cmss10 at 7pt
\def\IZ{\relax\ifmmode\mathchoice
   {\hbox{\cmss Z\kern-.4em Z}}{\hbox{\cmss Z\kern-.4em Z}}
   {\lower.9pt\hbox{\cmsss Z\kern-.4em Z}}
   {\lower1.2pt\hbox{\cmsss Z\kern-.4em Z}}\else{\cmss Z\kern-.4em Z}\fi}

\def\eqaligntwo#1{\null\,\vcenter{\openup\jot\m@th
\ialign{\strut\hfil
$\displaystyle{##}$&$\displaystyle{{}##}$&$\displaystyle{{}##}$\hfil
\crcr#1\crcr}}\,}
\catcode`\@=12

\def\oh{{1 \over 2}}
\def\b{\beta}
\def\a{\alpha}

\def\half{{1\over 2}}
\def\d{\dagger}

\def\pa{\partial}

\def\dj{\hbox{d\kern-0.347em \vrule width 0.3em height 1.252ex depth
-1.21ex \kern 0.051em}}

\nopagenumbers
\vsize=8.9in
\hsize=6.5in

{\baselineskip=16pt
\line{\hfil PUPT-1427}
\line{\hfil IASSNS-HEP-93/59}
\line{\hfil {\tt hep-th/9311015}}
\line{\hfill November 1993}
}

\bigskip\bigskip
\title{{\bf Glueball spectrum in a (1+1)-dimensional model for  QCD}}
\bigskip

\centerline {{\caps Kre\v simir Demeterfi}\foot{{\rm On leave of absence
from the Ru\dj er Bo\v skovi\'c Institute, Zagreb, Croatia}}
and {\caps Igor R. Klebanov} }
\centerline{\sl Joseph Henry Laboratories}
\centerline{\sl Princeton University}
\centerline{\sl Princeton, New Jersey 08544}
\bigskip
\centerline{{\caps Gyan Bhanot}}
\medskip
\centerline{
\vbox{\hsize2.7in
\centerline{\sl School of Natural Sciences}
\centerline{\sl Institute for Advanced Study}
\centerline{\sl Princeton, New Jersey 08540} }
\vbox{\hsize0.4in
\centerline{}
\centerline{and}
\centerline{}}
\vbox{\hsize2.7in
\centerline{\sl Thinking Machines Corporation}
\centerline{\sl 245 First Street}
\centerline{\sl Cambridge, MA 02142}}
}
\vskip 1.0in
\centerline{ABSTRACT}
\medskip
We consider (1+1)-dimensional QCD coupled to scalars in the
adjoint representation of the gauge group SU($N$).
This model results from dimensional reduction of the (2+1)-dimensional
pure glue theory. In the large-$N$ limit
we study the spectrum of glueballs
numerically, using the discretized \lcq. We find a discrete
spectrum of bound states, with the density of levels
growing approximately exponentially  with the mass.
A few low-lying states are very close to being
eigenstates of the parton number, and their masses can be
accurately calculated by truncated diagonalizations.
\vfill
\endpage

\pagenumbers

\centerline{\bf 1. INTRODUCTION}
\bigskip

The light-cone quantization supplemented with a regulator in the form of
discretized longitudinal momenta is a promising approach to
nonperturbative QCD [\Stan,\Brod,\Thorn], which may provide
tools for calculating the
hadron spectrum as well as the interaction cross-sections starting from first
principles. It has been successfully used in studies of
various field theories in 1+1 dimensions, and some progress
has been made in generalizing it to (3+1)-dimensional theories
[\Brod,\Tang].
Recently, Dalley and one of the authors applied this method to a new type of
models, namely, two-dimensional large-$N$ QCD coupled to matter
in the adjoint representation of SU($N$) [\DK]. These models are
far more complex than the large-$N$ QCD coupled to matter in the fundamental
representation [\GH--\FSmore], but they are the simplest models
where one can study some genuine QCD effects.
There are two color flux tubes attached to each quantum
and in that sense the quanta of the adjoint matter resemble physical
gluons.
The glueball-like bound states may contain any number of quanta connected into
a closed string by the color flux tubes.
Pair creation inside the string is not suppressed in the large-$N$
limit, leading to eigenstates which are  mixtures of strings with
different numbers of partons.
(In the 't Hooft model, due to the absence of transverse gluons,
one finds only the quark--antiquark bound states [\GH].)
Therefore, the adjoint matter imitates some transverse gluon
effects.

Another way to see this is to dimensionally reduce (2+1)-dimensional
pure glue theory,  and to note that the zero
mode of the transverse gluon field acts as the adjoint matter field
coupled to (1+1)-dimensional QCD.
Consider the pure glue SU($N$) theory in 2+1 dimensions,
$$ S=-{1\over 4g_3^2}\int d^3 x \Tr F_{\mu\nu}F^{\mu\nu}\,\,.\eqn\eq$$
Compactifying one of the spatial dimensions,
$y\sim y + L$, we may ignore the dependence of
fields on $y$  as $L\to 0$, i.e. $\partial A^\mu/\partial y =0$.
The action then reduces to
$$ S_{\rm sc} = \int d x^0 dx^1 \Tr \left[ \oh D_{\a}\F D^{\a}\F
-{1\over 4 g^{2}} F_{\a \b}F^{\a \b}\right], \eqn\action$$
where $g^2=g_3^2/L$, and $\phi (x^0, x^1)=A_y/g$
is a traceless $N\times N$ hermitian matrix field, whose
covariant derivative is given by $D_{\a}\phi = \pa_{\a}\phi
+i[A_{\a}, \phi]$. Therefore, $\phi$ represents the remnants of the
transverse gluon degrees of freedom.
In order to absorb the logarithmically divergent mass renormalization
it is necessary to add a mass term for $\phi$:
$$ S_{\rm sc} = \int d x^0 dx^1
 \Tr \left[ \oh D_{\a}\F D^{\a}\F +{1\over 2} m_0^2\F^2
-{1\over 4 g^{2}} F_{\a \b}F^{\a \b}\right]\,\,. \eqn\maction$$

The \lcq\ and the spectrum of the theory \maction\ were considered in
ref.~[\DK].
In this paper we present results of a new numerical diagonalization.
We find that holding
the renormalized mass fixed results in a convergent and
physically reasonable spectrum of glueballs. We compute this spectrum
by solving the
linear light-cone Schr\"odinger equation describing the bound states.
The density of
states grows roughly exponentially with their mass, as expected
in a physically interesting gauge theory. We also find that
a few low-lying states are very close to being eigenstates of the
parton number, while a typical excited state is a
complex mixture of states with
different parton numbers. These results are similar to the results
of our previous work [\BDK] where we performed a detailed numerical
study of a similar model, with the adjoint scalar replaced by an adjoint
Majorana fermion.  For a different approach to the model with
the adjoint fermions see ref.~[\dkut].

\bigskip
\centerline{\bf 2. LIGHT-CONE QUANTIZATION}
\bigskip

In this section we summarize the light-cone quantization of
$(1+1)$-dimensional SU($N$) gauge theory coupled to a scalar field
in the adjoint
representation [\DK].
Consider the action \maction\ and choose the \lc\ gauge $A_-=0$.
Introducing the light-cone coordinates $x^{\pm}=(x^0\pm x^1)/\sqrt 2$,
we obtain
$$S_{\rm sc}=\int dx^{+}dx^{-} \Tr \left[\pa_{+}\F\pa_{-}\F -\half m_0^2\F^2
+{1\over 2 g^2} (\pa_{-}A_{+})^2 + A_{+}J^{+} \right]\,\,, \eqn\mact$$
where
$J^{+}_{ij} = i[\F,\pa_{-}\F]_{ij}$ is the longitudinal momentum current.
A similar model, with $\phi$ taken to be a general complex matrix,
was considered in ref.~[\BPR] in an attempt to solve higher-dimensional
pure glue theory.

In the light-cone quantization $x^+$ is treated as the time and the canonical
commutation relations are imposed at equal $x^+$:
$$[\F_{ij}(x^{-}),\pa_{-}\F_{kl}(\tilde{x}^{-})] = {1\over 2}\,\,
\delta(x^{-}-\tilde{x}^{-})\delta_{il} \delta_{jk}\,\,.\eqn\ccr$$
The action does not depend on the
time derivatives of the gauge potential $A_{+}$ which,
therefore, can be eliminated by its constraint equations.
The \lc\ momentum and energy,  $P^{\pm}= \int dx^- T^{+\pm}$, are
found to be [\DK]:
$$ \eqalign{&P^{+} =  \int dx^{-} \Tr
[(\partial_{-}\F)^{2} ]\,\, ,\cr
&P^{-} = \int dx^{-} \Tr \left[\half m_0^2 \F^2
-  \half g^2  J^{+} {1\over \pa_{-}^{2}} J^{+}
\right]\,\, .\cr }\eqn\pminus$$
Our goal is to solve the eigenvalue problem
$$ 2P^+ P^- |\Phi\rangle= M^2 |\Phi\rangle\,\,, \eqn\ev$$
where the physical states must satisfy the zero-charge constraint
$$\int dx^- J^+  |\Phi\rangle = 0\ , \eqn\zc$$
arising from integration over the zero-mode of $A_+$.
Since $[P^+, P^-]=0$, $|\Phi\rangle$ is a simultaneous eigenstate
of $P^+$ and $P^-$. In practice it is easy to ensure that $|\Phi\rangle$
carries a definite $P^+$, but the subsequent solution of eq.~\ev\ is
highly non-trivial.

Introducing the mode expansion
$$\F_{ij}={1 \over \sqrt{2\pi}} \int_{0}^{\infty} {dk^{+}
\over \sqrt{2k^{+}}} \left(a_{ij}(k^{+}){\rm e}^{-ik^{+}x^{-}} +
a_{ji}^{\d}(k^{+}) {\rm e}^{ik^{+}x^{-}}\right)\,\,, \eqn\Four$$
with
$$ [a_{ij}(k^{+}),a_{lk}^{\d}(\tilde{k}^{+})] =
\delta(k^{+} - \tilde{k}^{+})
\delta_{il} \delta_{jk} \,\,,\eqn\modeccr$$
eq. \pminus\ can be written in terms of the oscillators as follows:
$$P^+ = \int_{0}^{\infty} dk\ k a_{ij}^{\d}(k)a_{ij}(k)\,\,, \eqn\nomess$$

$$\eqalign{&P^{-} =  \half m^2\int_{0}^{\infty} {dk\over k}\, a_{ij}^{\d}(k)
a_{ij}(k) \cr
&+ {g^2\over 8\pi}
 \int_{0}^{\infty} {dk_{1} dk_{2} dk_{3} dk_{4} \over
\sqrt{k_{1}k_{2}k_{3}k_{4}}} \biggl\{ A \D (k_{1}+k_{2}-k_{3}-k_{4})
a_{kj}^{\d}(k_{3})a_{ji}^{\d}(k_{4})a_{kl}(k_{1})a_{li}(k_{2})  \cr
& + B \D(k_{1} + k_{2} +k_{3} -k_{4})
\bigl(a_{kj}^{\d}(k_{1})a_{jl}^{\d}(k_{2})
a_{li}^{\d}(k_{3})a_{ki}(k_{4}) + a_{kj}^{\d}(k_{4})a_{kl}(k_{1})a_{li}(k_{2})
a_{ij}(k_{3})\bigr) \biggr\}\,\,, \cr }\eqn\mess$$
where
$$\eqalign{& A=
{(k_{2}-k_{1})(k_{4}-k_{3}) \over (k_{1} + k_{2})^2 } -
{(k_{3} + k_{1})(k_{4} + k_{2}) \over (k_{4}-k_{2})^2 }\,\,,\cr
\noalign{\vskip 0.2cm}
&B=  {(k_{1}+k_{4})(k_{3}-k_{2}) \over (k_{3}+k_{2})^2 } +
{(k_{3}+k_{4})(k_{1}-k_{2}) \over (k_{1}+k_{2})^2 }\,\, ,\cr}\eqn\coefbos$$
and we have dropped the superscript plus on $k_i$ for brevity.
The renormalized mass $m$ is defined by
$$ m^2= m_0^2 + {g^2 N\over 4\pi}\int_0^\infty {dp\over p}+
{g^2 N\over \pi}\int_{0}^{k} dp \,\,{k\over (k-p)^2 }\,\,.\eqn\eq
$$
Note that the second term in the mass renormalization is finite in the
principal value sense, while the first term is logarithmically
divergent. We will see that the spectrum is finite if the renormalized
mass $m$ is held fixed.

An important simplification in the \lcq\ is that the oscillator vacuum
satisfies
$$ P^+ |0\rangle=0\,;\qquad\qquad P^- |0\rangle=0\ .\eqn\eq$$
Other Fock states are then constructed by acting with creation
operators $a^\dagger_{ij}$ on the vacuum. The zero-charge condition
\zc\ requires that all the color indices be contracted.
Therefore, we look for eigenstates of eq.~\ev\ in the form
$$\eqalign{|\Phi(P^+) \rangle
=&\sum_{j=1}^\infty \int_0^{P^+} dk_1 \ldots dk_{j} \,
\delta\Bigl(\sum_{i=1}^{j} k_i-P^+\Bigr) \cr &
f_{j} (k_1, k_2, \ldots, k_{j})
N^{-j/2} \Tr \,[a^{\d}(k_1)\ldots a^{\d}(k_{j})] |0 \rangle\ , \cr }
\eqn\bw$$
where the wave functions have cyclic symmetry
$$f_i (k_2, k_3, \ldots, k_i, k_1)= f_i (k_1, k_2,k_3,\ldots, k_i)\,\,.
\eqn\cyclic$$
This state is trivially an eigenstate of $P^+$, and the problem is
to ensure that it is an eigenstate of $P^-$.

The complexity of
the coupling to adjoint matter arises mainly from the presence of pair
production and pair annihilation terms in $P^-$.
These terms appear in the leading order of the $1/N$ expansion,
provided that $g^2 N$ is kept fixed in the large-$N$ limit.
On the other hand, the terms in
$P^-$ that take one color singlet into two are suppressed by
$1/N$, which implies the stability of bound states in the large-$N$
limit. The glueball wave functions satisfy linear eigenvalue
equations [\KS, \DK], which,
upon introducing longitudinal momentum fractions $x_i=k_i^+/P^+$,
can be written as
$$\eqalign{& {M^2\pi\over g^2 N} f_i (x_1, x_2, \ldots, x_i)=
{m^2\pi\over g^2 N}\,{1 \over x_1}\,f_i(x_1,x_2,\ldots,x_i)
+\,{\pi\over 4\sqrt{x_1 x_2}}\,f_i(x_1,x_2,\ldots,x_i)
\cr \noalign{\vskip 0.2cm}
&+ \int_0^{x_1+x_2} dy\, {f_i (y, x_1+ x_2-y, x_3, \ldots, x_i)
\over 4\sqrt{x_1 x_2 y(x_1+x_2-y)}}\,
{(x_2-x_1)(x_1+x_2-2y)\over (x_1+x_2)^2}
\cr \noalign{\vskip 0.2cm}
&+ \int_0^{x_1+x_2} {dy\over (x_1-y)^2}\,
{(x_1+y)(x_1+2x_2-y)\over 4\sqrt{x_1 x_2 y(x_1+x_2-y)}}\,
\left[f_i(x_1,x_2,\ldots,x_i)-f_i(y,x_1+x_2-y,x_3,\ldots,x_i)\right]
\cr \noalign{\vskip 0.2cm}
&+\int_0^{x_1} dy \int_0^{x_1-y} dz
{f_{i+2} (y,z, x_1-y-z, x_2, \ldots, x_i)\over
4\sqrt{x_1 yz(x_1-y-z)}}
\left [{(x_1+y)(x_1-y-2z)\over (x_1-y)^2}+
{(2x_1-y-z)(y-z)\over (y+z)^2}\right ]\cr
\noalign{\vskip 0.2cm}
&+ {f_{i-2} (x_1+ x_2+x_3, x_4, \ldots, x_i)\over
4\sqrt{x_1 x_2 x_3(x_1+x_2+x_3)}}\,
\Bigl [{(2x_1+x_2+x_3)(x_3-x_2)\over (x_3+x_2)^2}+
{(x_1+x_2+2x_3)(x_1-x_2)\over (x_1+x_2)^2}\Bigr]\cr
\noalign{\vskip 0.3cm}
& \pm {\rm cyclic~ permutations~ of~} (x_1, x_2, \ldots, x_i)\,\,, \cr
}\eqn\RLS$$
where $M^2=2P^+ P^-$ is the bound state (mass)$^2$.
In writing this equation we have used a principal value integral [\FS]
$$\int_0^{x_1+x_2} {dy\over (x_1-y)^2}\,\,
{(x_1+y)(x_1+2x_2-y)\over \sqrt{y(x_1+x_2-y)}}
=-\pi\,\,.\eqn\reg$$
Note that the only singular term in eq. \RLS\ is well-defined in the
principal value sense.

By truncating eq. \RLS\ to the two-body sector [see eq.~(23)],
we can show using a constant variational wave function
that $M^2$ is unbounded from
below for $m^2<0$. For $m^2\geq 0$, on the other hand, the $M^2$ of
every state is positive. Thus, the critical value of
$m^2$ is zero. We believe that this is the correct value to consider if
we want to regard our model as the proper dimensional reduction of the
(2+1)--dimensional pure glue QCD. Therefore, we will be interested
primarily in the $m^2=0$ case.

The form of eq. \RLS\ indicates that the interactions conserve the
number of partons modulo 2. Therefore, there are two kinds of
bound states: those involving mixtures of states with even parton
numbers, and those involving mixtures of states with odd parton
numbers. From here on, we will call them even and odd glueballs
respectively. Furthermore,
eq. \RLS\ possesses another $\IZ_2$ symmetry, $T$ [\dkut]:
$$f_i (x_1, x_2, \ldots, x_i)=T f_i (x_i, \ldots, x_2, x_1)\,\,.\eqn\eq $$
The $\IZ_2$ quantum number
$T$ has two possible values, 1 and $-1$. In terms of the original
field, $T: \phi_{ij} \to \phi_{ji}$, which obviously leaves $P^\pm$
invariant [\dkut]. Physically, every bound state can be thought of as a
superposition of
oriented closed strings, and the quantum number $T$ describes the
transformation property under a reversal of orientation.

\bigskip
\centerline{\bf 3. THE DISCRETIZED APPROXIMATION AND NUMERICAL
RESULTS}
\bigskip

The system of equations \RLS\ involves an infinite number of
multivariable functions.
It seems very difficult, if not impossible, to solve this system of
equations analytically.
Therefore, here we follow the
same strategy as in [\BDK] and replace the
continuum
equations \RLS\ by a sequence of discretized approximations, such that
the eigenvalues of the discretized problems eventually converge to the
eigenvalues of \RLS.
In the \lc\ quantization, a simple discretized approximation is
obtained by replacing
the continuous momentum fractions $x$ by a discrete set $n/K$,
where $n$ are odd positive integers, and
the positive integer $K$ is sent to infinity as the cut-off
is removed [\Thorn,\Brod].
The restriction to odd integers corresponds to the choice of antiperiodic
boundary conditions in the discretized light-cone quantization,
and leads to the best convergence towards continuum limit.
Thus, the functions $f_i (x_1, x_2, \ldots, x_i)$ are
replaced by
$$ g_i (n_1, n_2, \ldots, n_i)= K^{(1-i)/2}
f_i\,\Bigl({n_1\over K},{n_2\over K},\ldots,{n_i\over K}\Bigr)$$
with $\sum_{j=1}^i n_j=K$, and
$$\int_0^1 dx \to {2\over K}\sum_{{\rm odd~}n>0}^K\,\,.  $$
For the even glueballs,
$g_i (n_1, n_2, \ldots, n_i)$ are normalized as
$$\sum_{{\rm even~}i}  i \sum_{{\rm odd~}n_1>0}\ldots
\sum_{{\rm odd~}n_{i-1}>0}
\Bigl|g_i (n_1, n_2, \ldots, n_{i-1}, K- \sum_{j=1}^{i-1} n_j)
\Bigr|^2 =1 \ ,
\eqn\conorm$$
and similarly for the odd glueballs.

The number of independent components of
$ g_i (n_1, n_2, \ldots, n_i)$
is equal to the number of partitions of
$K$ into positive odd integers, modulo cyclic permutations.
Finding all such partitions
is a combinatorial problem that is easily solved with a computer
program.
If we regard all the independent
$ g_i (n_1, n_2, \ldots, n_i)$ as components of a vector, then
the discretized eigenvalue problem \RLS\ is equivalent to diagonalizing
a matrix
$${M^2\pi\over g^2 N} = K\left(x V +T\right) \ ,\eqn\fham$$
where $x={\pi m^2\over g^2 N}$ is
the dimensionless parameter.
The main difficulty in this approach is that  the number of states
increases rapidly with $K$. For example, there are
765, 1169, 1810, 2786, 4340, 6712 states for
$K=20, 21, 22, 23, 24, 25$, respectively.
In actual calculations it is advantageous to consider separately
the even and odd sectors under $T$, and introduce the linear
combinations
$$g_i^{\pm} (n_1, n_2, \ldots, n_i) =
{1\over \sqrt{2}}\,\Bigl(g_i (n_1, n_2, \ldots, n_i) \pm
g_i (n_i, n_{i-1}, \ldots, n_1) \Bigr)\,\,.$$
The entries of $V$ and $T$ can be read off from eqs. \RLS.
For example, for $K=6$ there are 4 partitions which correspond to
the $\IZ_2$-even sector: $\{(1,1,1,1,1,1), (3,1,1,1), (5,1),
(3,3)\}$.  The matrix to be diagonalized in this case reads:
$$ K(xV+T)=\left(\matrix
{ 36x + 9\pi & 0 & 0 & 0 \cr
0 & 20x + (3+\sqrt{3})\pi & 0 & 0 \cr
0 & 0 & {36x\over 5} + {16+3\pi\over\sqrt{5}} & -\sqrt{{512\over 5}} \cr
0 & 0 &  -\sqrt{{512\over 5}} & 4x + {32\over\sqrt{5}}+\pi }
\right) \ .$$
For this simple example there are no partitions corresponding to the
$\IZ_2$-odd sector. For larger values of $K$, however, the total
number of partitions splits approximately equally between the even
and odd sectors, allowing us to reach higher values of $K$.
For $K=24$ there are 2358 components in the $\IZ_2$-even sector
and 1982 components in the $\IZ_2$-odd sector. For
$K=25$ there are 3544 and 3168 even and odd components, respectively.

A good numerical procedure is to calculate the spectrum for
a fixed
$x$ and a range of values of $K$, and then to extrapolate the results to
infinite $K$, the continuum limit. We will also assume that some bulk
properties of the spectrum can be estimated from the results at a fixed
large $K$. We will be most interested in
$x=0$ which corresponds to the limit of massless quanta.
Note that, with our choice of the regulator, we observe only
even (odd) glueballs
for an even (odd) $K$.
We will present here the results for our
biggest diagonalization in each sector separately: for even $K=24$
and for odd  $K=25$.
Our analysis is very similar to the one performed in ref.~[\BDK] for the
model with the adjoint fermions.

In fig.~1 we show the spectrum of even glueballs
for $x=0$ and $K=24$,
with the mass plotted vs. the expectation value of the number of
partons, $n$. The density of states obviously increases rapidly with
the mass, and almost all the states lie within a band bounded
by two $\langle n \rangle \sim M$ lines.
One interesting feature of our results, already noted in ref.~[\DK]
for smaller $K$, is that for a few low-lying eigenstates the wave
functions are strongly peaked on states with a definite number of
partons. For example, for $K=24$ the ground state has probability
0.99899 to consist of two partons, and the first excited state has
probability 0.99734 to consist of four partons.
For the ground state, it is a good approximation, therefore, to
truncate eq. \RLS\ to the two-body sector described by the
wave function $\phi(x)= f_2 (x, 1-x)$. From eq. \cyclic\
we have $\phi(x)= \phi(1-x)$. Thus, we obtain
$$ M^2 \phi(x) = m^2 {\phi(x)\over x(1-x)}
+{g^2 N\over 2} {\phi(x)\over \sqrt{x(1-x)}}+
{g^2 N\over 2\pi}
\int_0^1 dy{\phi(x)-\phi(y)\over (x-y)^2}\,
{(x+y)(2-x-y)\over \sqrt{x(1-x) y(1-y)}}\,\,.
\eqn\trunc$$
This is simply the bound state
equation for the theory with scalar quarks in the
fundamental representation of SU($N$) [\FS,\FSmore], with $g^2\to 2g^2$.
The interaction strength is doubled because now there are two flux tubes
connecting the pair of partons. The lowest eigenvalue of eq. \trunc,
$M^2=4.76 g^2N/\pi$,
provides a good upper bound on the lowest eigenvalue of eq. \RLS.

As the excitation number
increases, however, the wave functions typically become quantum
superpositions of states with different parton numbers.
The physical picture is that a typical excited state contains
some number of virtual pairs, and our data supports this expectation.
In order to quantify this effect, we will call a state pure if it
has probability $> 0.9$ to be in one of the number sectors.
Table 1 shows the total number of states and the number of pure states
in each mass interval of fig.~1. We also show
the expectation value of the number of partons averaged over all states
in each mass interval.
Evidently, a few low-lying states are pure, while there are almost
no pure states among the high excitations.
In fig.~2 we plot the spectrum of the odd glueballs for $x=0$ and $K=25$, and
in table 2 we quantify their purity.
The lightest odd glueball contains, to a good approximation,
only three partons, and the first
excited state is almost purely a five-parton state.
If we increase the mass of the quantum, the pair creation
becomes somewhat suppressed. We have performed the diagonalization
for $x=1$ and found, indeed, that the number of pure states increases.
While the distributions of states
for $x=0$ shown in figs.~1 and 2 are almost uniform within a band,
for $x=1$ there is an increase in the number
of states with the average parton number
near $2,4,6,8,\ldots$ for $K=24$, and a similar increase
in the number of states with the average parton
number near $3,5,7,9,\ldots$ for $K=25$.

A striking property of figs.~1 and 2 is the rapid growth of the
density of states
with increasing mass. In fig.~3 we plot the logarithm of the number of
odd glueballs vs. the mass for the data in table 2. For a certain range of
masses the graph is approximately linear. The deviation from linearity
for large enough mass is clearly due to the effects of the cut-off.
Our results indicate that the density of states grows roughly
exponentially with the mass, exhibiting the Hagedorn behavior
$$\rho(m)\sim m^\alpha e^{\beta m}\,\,.
\eqn\Hag$$
Thus, although the mass spectrum is discrete, it rapidly becomes
virtually indistinguishable from a continuum. From our data we estimate
that the inverse Hagedorn temperature is
$\beta\approx (0.65 - 0.7) \,\sqrt{\pi/ (g^2 N)}$.

Another physical effect that is pronounced in our results
(tables 1 and 2) is that
the mass increases roughly linearly with the average number of partons.
In fig.~4  we plot these results for $x=0$ and $K=24$ (table 1).
A heuristic explanation
of this effect was suggested in [\BDK] using the result of ref.~[\KKS]
that the ground state energy of a system of $n$ non-relativistic
particles connected into a closed string by harmonic springs,
which should be identified with $M^2$, behaves as $\sim n^2$ for
sufficiently large $n$.

Since the low-lying states are very pure, they can be well approximated
by truncating the diagonalization to a single parton number sector.
For instance, for $x=0$ and $K=24$ the ground state has probability
0.999999 to consist of 2, 4 or 6 partons.
Extrapolating to infinite $K$, we find 0.99996, and therefore
this truncation is highly reliable.
In table 3 we compare the full and truncated calculations.
We have performed the truncated diagonalizations
up to $K=40$, and
extrapolating these results  to infinite $K$,
we find the upper bounds $M^2\approx 4.33 \,g^2 N / \pi$
for $x=0$ and $M^2\approx 13.43 \,g^2 N / \pi$
for $x=1$. These are extremely close to the extrapolations from
exact diagonalizations. This shows that, by judiciously truncating the space
of states, certain eigenvalues can be determined to a good accuracy with
relatively small diagonalizations.
Similar approximations work well for the odd glueballs.
For $x=0$ and $K=25$ the lightest
state has probability 0.999992 to consist
of three or five partons. Extrapolating it
to infinite $K$, we find  0.99994 .
The advantage of this truncation is that we
can access higher value of $K$ (up to 51) than in the full
diagonalization, and extrapolate more reliably.
We find that the lowest eigenvalue extrapolated
in this fashion is
$M^2\approx 11.95 \,g^2 N / \pi$ for $x=0$, and
$M^2\approx 30.87 \,g^2 N / \pi$ for $x=1$.
Good agreement of these values with extrapolations from exact
diagonalizations gives us some confidence that our methods are consistent.
We have used the Bulirsch-Stoer algorithm which has proved to
be particularly efficient for extrapolating short series [\ext].
In fig.~5 we show the $M^2$ of the ground state
for $x=0$ as a function of $1/K$.
For $K\le 24$ we show the results of the full
diagonalization, while for $26 \le K \le 40$ we use the results
of truncated diagonalization in the $2-, 4-$ and $6-$bit sector.

\bigskip
\centerline {\bf 4. DISCUSSION}
\bigskip

Our calculations suggest that
(1+1)-dimensional large-$N$ QCD coupled to adjoint scalar matter
captures some of the physical features of the
higher-dimensional pure glue gauge theories. This model may also serve as
a good test of the \lcq\ methods that are promising to become
a useful tool for studying the non-perturbative structure of the strong
interactions. We
mention below some interesting questions for further study.

1. Even though the model is (1+1)-dimensional, we have observed
a rich structure of glueball states. In fact, it is
much richer than in
the models coupled to matter in the fundamental
representation of SU($N$), which have only one state per unit mass-squared
[\qcd, \FS].
It would be interesting to
gain further insight into the Hagedorn behavior, eq. \Hag.
Could there be exact formulae for the constants $\alpha$ and
$\beta$? Their determination is relevant to the properties of
the deconfining phase transition. Such a transition
 has been recently studied in the
model with adjoint fermions [\dkut, \kog].

2. We found that some low-lying states are exceedingly close
to being eigenstates of the parton number. However, they fail
to be exact eigenstates. We feel that this deserves a deeper
understanding. If this property holds true in higher-dimensional
gauge theories, it could lead to an enormously simplified picture
of the low-lying states.

3. The glueball states can be simply pictured as closed strings
that cannot split or join in the large-$N$ limit. Is there
a continuum string description of our model?
A study of this question seems to be a logical next step in
the program of ref. [\DG], where a string representation
of the pure glue theory was found.

\ack

We are grateful to C. Callan, S. Dalley,
D. Gross, S. Gubser, A. Polyakov, S. Shenker, and
especially D. Kutasov, for
helpful discussions.
We also thank  Thinking Machines Corporation for use of their
computers.
I. R. K. and K. D. were supported in part by
NSF Presidential Young Investigator program under Grant No. PHY-9157482 and
James S. McDonnell Foundation Grant No. 91-48. I. R. K. was also
supported by DOE Grant No. DE-AC02-76WRO3072 and
the A. P. Sloan Foundation.
The work of G. B. was partly supported by a US DOE Grant No.
DE-FG02-90ER40542 and by the Ambrose Monell Foundation.

\refout
\endpage

\noindent Table 1. Numerical data for $K=24$, $x=0$ shown in fig.~1;
the 3.0 bin includes all states whose masses are $1.5-3.0$; etc.
\medskip
\tablewidth=6.5in
\begintable
$M$  \| 3.0 | 4.5 | 6.0 |7.5 | 9.0 | 10.5 | 12.0 | 13.5 | 15.0 | 16.5\cr
number of states \| 1|1|1|2|7|26|103|262|598|1013 \cr
number of pure states\| 1|1|1|1|2|3|4|3|1|1\cr
average parton number \| 2.00|4.00|2.17|4.98|4.64|4.96|6.01|
7.36|8.54|9.91
\endtable

\bigskip\bigskip\bigskip

\noindent Table 2. Numerical data for $K=25$, $x=0$ shown in fig.~2.
\medskip
\tablewidth=6.5in
\begintable
$M$  \| 4.5| 6.0 |7.5 | 9.0 | 10.5 | 12.0 | 13.5 | 15.0|16.5|18.0 \cr
number of states \|1|1|1|7|26|99|277|651|1270|1757 \cr
number of pure states\| 1|1|1|3|5|5|1|0|0|0\cr
average parton number \| 3.00|4.99|3.09|4.52|5.12|5.96|7.18|8.39|
9.69|11.12
\endtable

\endpage

\noindent Table 3. Numerical values of the ground state mass for
$x=0$ and $x=1$ for even $K$.
\medskip
\tablewidth=6.5in
\begintable
\hfil\|\multispan{2}\tstrut\hfil $x=0$ \hfil \|
       \multispan{2}\tstrut\hfil $x=1$ \hfil \crthick
$K$  \| $M^2$ (full) | $M^2$ (2+4+6--bit) \|
$M^2$ (full) | $M^2$ (2+4+6--bit) \crthick
12\|4.1064|4.1064\|10.8712 |10.8712 \cr
14\|4.1476|4.1476\|11.0652 |11.0652 \cr
16\|4.1789|4.1789\|11.2219 |11.2219 \cr
18\|4.2034|4.2034\|11.3520 |11.3520 \cr
20\|4.2230|4.2230\|11.4624 |11.4624 \cr
22\|4.2388|4.2388\|11.5577 |11.5577 \cr
24\|4.2518|4.2519\|11.6411 |11.6411 \cr
$\vdots$\|$\vdots$| $\vdots$ \| $\vdots$ | $\vdots$\cr
$\infty$ \|4.3 |4.33\|13.4 |13.43 \endtable

\bigskip

\noindent Table 4. Numerical values of the ground state mass for
$x=0$ and $x=1$ for odd $K$.
\medskip
\tablewidth=6.5in
\begintable
\hfil\|\multispan{2}\tstrut\hfil $x=0$ \hfil \|
       \multispan{2}\tstrut\hfil $x=1$ \hfil \crthick
$K$  \| $M^2$ (full) | $M^2$ (3+5--bit) \|
$M^2$ (full) |  $M^2$ (3+5--bit) \crthick
15 \| 10.5001 | 10.5003  \|26.0496  |26.0497 \cr
17 \| 10.6569 | 10.6572  \|26.5323  |26.5326 \cr
19 \| 10.7853 | 10.7858  \|26.9358  |26.9362 \cr
21 \| 10.8925 | 10.8932  \|27.2797  |27.2802 \cr
23 \| 10.9835 | 10.9843  \|27.5775  |27.5781 \cr
25 \| 11.0616 | 11.0626  \|27.8387  |27.8393 \cr
$\vdots$\|$\vdots$|$\vdots$\| $\vdots$ | $\vdots$\cr
$\infty$ \|11.9 |11.95 \|30.8|30.87 \endtable

\endpage

\centerline{FIGURE CAPTIONS}
\bigskip

\item
{\bf Fig.1.} The spectrum of even glueballs for $K=24$,
$x=0$, $M <18$; mass $M$  is measured in units of
$\sqrt{g^2 N / \pi}$ and plotted vs. the expectation value
of the parton number.
\item
{\bf Fig.2.} The spectrum of  odd glueballs for $K=25$,
$x=0$, $M < 18$.
\item
{\bf Fig.3.} Logarithm of the density of states and a linear fit
for $K=25$ and $x=0$.
\item
{\bf Fig.4.} Average number of partons as a function
of mass for $K=24$ and $x=0$.
\item
{\bf Fig.5.} Extrapolation towards infinite $K$ of the ground state
at $x=0$; $M^2\pi/g^2N$ is plotted versus $1/K$.
\endpage

\bye